\documentclass[twocolumn,showpacs,aps,pre]{revtex4}
\usepackage{makeidx}
\usepackage{amsmath}
\usepackage{amssymb}
\usepackage{graphicx}
\usepackage{graphics}

\begin{document}

\title{Surface segregation of conformationally asymmetric polymer
blends}
\author{Semjon Stepanow}
\author{Andrei A. Fedorenko}
\affiliation{Fachbereich Physik, Martin-Luther-Universit\"{a}t Halle, D-06099 Halle,
Germany}
\date{\today}

\begin{abstract}
We have generalized the Edwards' method of collective description of dense
polymer systems in terms of effective potentials to polymer blends in the
presence of a surface. With this method we have studied conformationally
asymmetric athermic polymer blends in the presence of a hard wall to the
first order in effective potentials. For polymers with the same gyration
radius $R_g$ but different statistical segment lengths $l_{A}$ and $l_{B}$
the excess concentration of stiffer polymers at the surface is derived as $%
\delta \rho _{A}(z=0)\sim (l_{B}^{-2}-l_{A}^{-2}){\ln (}R_{g}^{2}/l_{c}^{2}{)%
}$, where $l_{c}$ is a local length below of which the incompressibility of
the polymer blend is violated. For polymer blends differing only in degrees
of polymerization the shorter polymer enriches the wall.
\end{abstract}

\maketitle

%\email[Corresponding author: email\ ]{stepanow@physik.uni-halle.de}

\section{Introduction}

\label{sec1}

The effects of surfaces on the behavior of polymer melts and
blends is of basic importance in their numerous applications such
as adhesion, lubrication, wetting, catalysis, etc.
\cite{binder-actap95}. The structure and properties of the blends
and other polymeric materials within a few nanometers at a surface
can differ significantly from corresponding properties in the
bulk. For example, in polymer blend a segregation of one of the
components to the surface is possible even if the blend is
miscible in bulk. Hence the questions how the surface can change
the properties of polymeric materials and how it may be controlled
are of practical interest. Despite the large theoretical and
experimental interest in the behavior of polymer blends in the
presence of surfaces and achieved basic understanding there is no
satisfactory analytical treatment of segregation of polymers at
surfaces. Most studies of polymer blends near surfaces are based
on phenomenological expressions for a free energy, which include
surface terms that account for adsorption or repulsion of a
particular type of monomers \cite{binder,fredrickson87}.
Minimization of the free energy gives equilibrium concentration
profiles for each component. There exist more rigorous approaches,
which allow one to derive the concentration profiles starting from
the microscopic polymer statistics in the presence of a surface.
One of such microscopic approaches is the integral equations
method which can be applied to various site-site or hard-particle
models of a dense polymeric system \cite{yethiraj94}. This method
having many advantages such as ability to predict microscopic
correlations between different types of monomers and between
monomers and surfaces requires a considerable amount of numerical
computations. Most of analytically treatable methods, which rely
on the continuum Gaussian chain model, take into account the
monomer-monomer
interactions usually using either the random phase approximation \cite%
{leibler80}, which is most suited to treat systems in the weak segregation
limit or self-consistent-field theories \cite{helfand72,freed}, which are
most suited to treat systems in the strong segregation limit.

Recently there has been a big deal of attention on the surface segregation
due to conformational asymmetry of the molecules of the polymer blend and
due to differences in topology \cite%
{foster92,sikka93,fredrickson-donley,yethiraj94,yethirj95,kumar95,wu-fredrik-carton,walton-mayes96,donley-wu-fredrik,tretinnikov98,chapman05}%
. It was established that the composition of polymer blends in the vicinity
of surfaces can be different from the bulk composition even for neutral
surfaces. It was found that for polymer blends composed of polymers with
different degree of polymerization but of chemically identical monomers,
shorter polymers are in excess at the wall. It was also demonstrated in
simulations \cite{yethiraj94,yethirj95} and supported by calculations using
the integral equation theory \cite{yethiraj91} that stiffer polymers are
present in excess in the vicinity of the wall. However the self-consistent
field theory developed in Ref.~\cite{wu-fredrik-carton} predicts the
opposite effect, i.e., the excess of more flexible polymers. Unfortunately,
no predictions on the behavior of polymer blends of chemically identical
polymers with different degrees of polymerization were made in Ref.~\cite%
{wu-fredrik-carton}.

In this paper we present the analytical study of the behavior of athermic
polymer blend in the presence of a hard wall using the generalization of the
Edwards' collective description of dense polymer systems in terms of
effective potentials \cite{edwards65,edwards75} to polymer blends in the
presence of a neutral surface. The bare one-polymer Green's function $G$
obeys the Dirichlet boundary condition. We show that a partial summation of
graphs results in replacing the bare $G$ with the effective one-polymer
Green's function $G_{r}$, which, as we argue, obeys the reflecting boundary
condition. The bare and effective Green's functions are related by the Dyson
equation, where the self-energy $\Sigma $ is defined by series of graphs.
This part of our work is similar to Ref.~\cite{wu-fredrik-carton}, however
with the following significant difference. In the present work the Dyson
equation results in an integral equation for $G_{r}$, which determines the
relevant reference state to describe polymer melts in the presence of a
neutral wall. The concentration profiles are due to fluctuations, which are
not taken into account in self-consistent field theories. The method we use
can be applied in a straightforward way to study the behavior of polymer
blends and copolymer melt in the presence of selective surfaces, the
dimensions of polymer molecules in the melt, the distribution of polymer
ends, etc.

The paper is organized as follows. Section \ref{sec2a} outlines the statistics
of a single polymer chain in the presence of
a hard wall. Section \ref{sec2b} introduces to the collective description
of dense polymer system. Section \ref{sec2c} contains the discussion of
the behavior of the effective potentials and screening effects in the
presence of a hard wall. Subsection \ref{sec2d} introduces to the collective
description in the presence of a neutral surface. Section \ref{comput}
contains calculations of the excess monomer concentration of constituents of
an incompressible athermic polymer blend in the vicinity of a hard wall.
Section \ref{concl} contains our conclusions.

\section{Collective description of dense polymer systems}

\label{sec2}

\subsection{Polymer chains in the presence of a hard wall}

\label{sec2a}

The Green's function of a free polymer, which is proportional to the
relative number of configurations of the ideal chain with the ends fixed at $%
\mathbf{r}$ and $\mathbf{r}^{\prime }$, and gives under appropriate
normalization the distribution function of the end-to-end distance, obeys
the Schr\"{o}dinger type differential equation \cite{edwards65,degennes69}
\begin{equation}
\left[ \frac{\partial }{\partial N}-a^{2}\nabla _{\mathbf{r}}^{2}\right] G(%
\mathbf{r},N;\mathbf{r}^{\prime })=\delta (\mathbf{r}-\mathbf{r}^{\prime
})\delta (N),  \label{b3}
\end{equation}%
where $N$ is the number of statistical segments, and $a^{2}=l^{2}/6$ with $l$
being the statistical segment length of the chain. The distribution function
of the end-to-end distance obtained from Eq.~(\ref{b3}) reads
\begin{equation}
G_{0}(r,N;0)=\left( \frac{1}{4\pi a^{2}N}\right) ^{3/2}\exp \left( -\frac{%
r^{2}}{4a^{2}N}\right).  \label{bg}
\end{equation}

In the presence of a hard wall we have to impose an appropriate boundary
condition on Eq.~(\ref{b3}). For polymers in a dilute solution with a hard
wall situated at $z=0$ one should use the Dirichlet boundary condition%
\begin{equation}
G(\mathbf{r},N;\mathbf{r}^{\prime })|_{z=0}=0,  \label{dirichlet}
\end{equation}%
where $\mathbf{r}\equiv \left\{ \mathbf{r}_{\shortparallel },z\right\} $.
The solution of Eq.~(\ref{b3}) with the boundary condition (\ref{dirichlet})
is given by
\begin{eqnarray}
G(\mathbf{r},N;\mathbf{r}^{\prime }) &=&G_{0}(\mathbf{r}_{\shortparallel }-%
\mathbf{r}_{\shortparallel }^{\prime },N;0)\left[ G_{0}(z-z^{\prime
},N;0)\right.  \notag \\
&-&\left. G_{0}(z+z^{\prime },N;0)\right] .  \label{G_diri}
\end{eqnarray}

It was argued in Ref.~\cite{silberberg68} that for an incompressible polymer
melt in the presence of a neutral surface one should impose the reflecting
(Neumann) boundary condition%
\begin{equation}
\partial _{z}G(z,N;z^{\prime })|_{z=0}=0  \label{neum}
\end{equation}%
on the Green's function of single polymer chains. The solution of Eq.~(\ref%
{b3}) with the boundary condition (\ref{neum}) is given by
\begin{eqnarray}
G(\mathbf{r},N;\mathbf{r}^{\prime }) &=&G_{0}(\mathbf{r}_{\shortparallel }-%
\mathbf{r}_{\shortparallel }^{\prime },N;0)\left[ G_{0}(z-z^{\prime
},N;0)\right.  \notag \\
&+&\left. G_{0}(z+z^{\prime },N;0)\right] .  \label{b4}
\end{eqnarray}%
The Laplace transform of the $z$ part of the Green's function with respect
to $N$, which we will need in the following, is given by
\begin{equation}
G_{0}(z-z^{\prime },p)=\frac{1}{2a\sqrt{p}}\exp \left(-\mid z-z^{\prime }\mid
\frac{\sqrt{p}}{a}\right).  \label{b6}
\end{equation}

The monomer density of a single polymer chain%
\begin{equation}
n(z,N)=\int_{0}^{N}ds\langle \delta (z-z(s))\rangle  \label{nz}
\end{equation}%
can be expressed through the Green's function of the polymer chain as follows%
\begin{equation}
n(z,N)=\int_{0}^{N}ds\int\limits_{0}^{\infty }dz^{\prime
}\int\limits_{0}^{\infty }dz^{\prime \prime }G(z^{\prime
},z,N-s)G(z,z^{\prime \prime },s).  \label{nz-p}
\end{equation}%
The straightforward computation using the Green's function obeying adsorbing
and reflecting boundary condition yields%
\begin{eqnarray}
n(z,N) &=&N\left[ 2\mathrm{erf}({z}/{2})+{z}^{2}[1+\mathrm{erf}({z}/{2}%
)]\right.  \notag \\
&+&\frac{2z}{\sqrt{\pi }}\exp (-{z^{2}}/{4})-\mathrm{erf}({z})  \notag \\
&-&\left. 2z^{2}\mathrm{erf}({z})-\frac{2z}{\sqrt{\pi }}\exp (-{z^{2}})%
\right] ,  \label{rho4}
\end{eqnarray}%
\begin{equation}
n(z,N)=N,  \label{nz3}
\end{equation}%
respectively. The distance $z$ in Eq.~(\ref{rho4}) is measured in units of $%
R_{g} $. The monomer density of one chain in the presence of a hard wall
does not depend on the distance to the wall in the case of the reflecting
boundary condition. The multiplication of $n(z,N)$ in Eqs.~(\ref{rho4}) and (%
\ref{nz3}) with the number of chains per volume $n/\mathcal{V}$ gives the
monomer density of a mixture of independent chains. The necessity of change
of the boundary condition in the polymer melt will be discussed at the end
of Section~\ref{sec2d}.

\subsection{Collective description of the polymer mixture in bulk}

\label{sec2b}

In the analytical approach to the description of dense polymer systems due
to Leibler \cite{leibler80} the random phase approximation is used to derive
the Ginzburg-Landau type functional of the diblock copolymer melt as a
functional of the order parameter. The collective description of
concentrated polymer systems due to Edwards \cite{edwards75} gives the
physical quantities under interest as series in powers of the effective
potentials. These series are closely related to those in the theory of
polymer solutions \cite{descloizeaux} with the main difference that the bare
interaction potentials between the monomers are replaced by the effective
ones (see below). The diagrammatic way of introduction the collective
description in Ref.~\cite{stepanow95} enables one to go beyond the random
phase approximation and establishes the connection between Leibler's and
Edwards' approaches.

We now will consider the collective description of the polymer mixture
consisting of $A$ and $B$ polymers in terms of effective potentials
following Ref.~\cite{stepanow95}, where this approach was developed for
copolymer melt. The elastic part of the Edwards free energy of $n_{A}$
polymers of type $A$ and $n_{B}$ polymers of type $B$ chains confined to a
volume $V$ is given by
\begin{equation}
F_{\mathrm{el}}=\frac{3}{2l^{2}}\sum\limits_{m=1}^{n_{A}+n_{B}}%
\int_{0}^{N}ds\left( \frac{d\mathbf{r}_{m}(s)}{ds}\right) ^{2},  \label{b20}
\end{equation}%
where $\mathbf{r}_{m}(s)$ parametrizes the configuration of $m\mathrm{th}$
polymer as a function of the position along the chain $s$. The interaction
part of the free energy (in units of $k_{B}T$) of the blend can be written
using the microscopic monomer densities of both polymers
\begin{eqnarray}
\rho _{A}(\mathbf{r}) &=&\sum\limits_{m=1}^{n_{A}}\int_{0}^{N}ds\,\delta (%
\mathbf{r}-\mathbf{r}_{m}(s)),  \notag \\
\rho _{B}(\mathbf{r})
&=&\sum\limits_{m=n_{A}+1}^{n_{A}+n_{B}}\int_{0}^{N}ds\,\delta (\mathbf{r}-%
\mathbf{r}_{m}(s))  \label{b18}
\end{eqnarray}%
in the form%
\begin{equation}
F_{\mathrm{int}}=\frac{1}{2}\int d^{3}r_{1}\int d^{3}r_{2}\rho _{\alpha }(%
\mathbf{r}_{1})V_{\alpha \beta }(\mathbf{r}_{1}-\mathbf{r}_{2})\rho _{\beta
}(\mathbf{r}_{2}),  \label{b21}
\end{equation}%
where
\begin{equation}
V_{\alpha \beta }(\mathbf{r}_{1}-\mathbf{r}_{2})=\left(
\begin{array}{cc}
V & V+\chi \\
V+\chi & V%
\end{array}%
\right) \delta ^{(3)}(\mathbf{r}_{1}-\mathbf{r}_{2})  \label{b22}
\end{equation}%
($\alpha $,$\beta =A$, $B$) is the interaction matrix of monomer-monomer
interactions, and $\chi $ is connected to the Flory-Huggins parameter. The
sum convention over repeated indices is implied in Eq.~(\ref{b21}) and
henceforth.

Let us now start with the computation of the average concentration of one of
the polymers%
\begin{equation}
\left\langle \rho _{\alpha }(\mathbf{r})\right\rangle =\frac{\int
Dr_{i}(s)\rho _{\alpha }(\mathbf{r})\exp \left( -F_{\mathrm{el}}-F_{\mathrm{%
int}}\right) }{\int Dr_{i}(s)\exp \left( -F_{\mathrm{el}}-F_{\mathrm{int}%
}\right) }  \label{rho1}
\end{equation}%
using the collective description of the polymer blend in terms of effective
potentials. The average monomer density can be written after introducing a
two-component field $\Phi _{\alpha }(r)$ in the equivalent form as follows%
\begin{eqnarray}
&&\!\!\!\!\!\!\left\langle \rho _{\alpha }(\mathbf{r})\right\rangle  \notag
\\
&&\!\!\!=\frac{\int Dr_{i}(s)\int D\Phi (\mathbf{r})\prod_{\mathbf{r}%
^{\prime },\beta }\delta (\Phi _{\beta }-\rho _{\beta })\Phi _{\alpha }(%
\mathbf{r})e^{-F_{\mathrm{el}}-F_{\mathrm{int}}}}{\int Dr_{i}(s)\int D\Phi (%
\mathbf{r})\prod_{\mathbf{r}^{\prime },\beta }\delta (\Phi _{\beta }-\rho
_{\beta })e^{-F_{\mathrm{el}}-F_{\mathrm{int}}}}.  \notag \\
&&  \label{rho2}
\end{eqnarray}%
The insertion of the Fourier transformation of the infinite product of
$\delta$-functions
\begin{equation*}
\prod_{\mathbf{r}^{\prime },\beta }\delta (\Phi _{\beta }(\mathbf{r}^{\prime
})-\rho _{\beta }(\mathbf{r}^{\prime }))=\int DQ(\mathbf{r})e^{iQ\cdot (\Phi
-\rho )}
\end{equation*}%
%
%
%%%%%%%%%%%%%%%%%%%%%%%%%%%%%%%   Figure 1  %%%%%%%%%%%%%%%%%%%%%%%%%%%%%%%%
\begin{figure}[tbph]
\includegraphics[clip,scale=0.45]{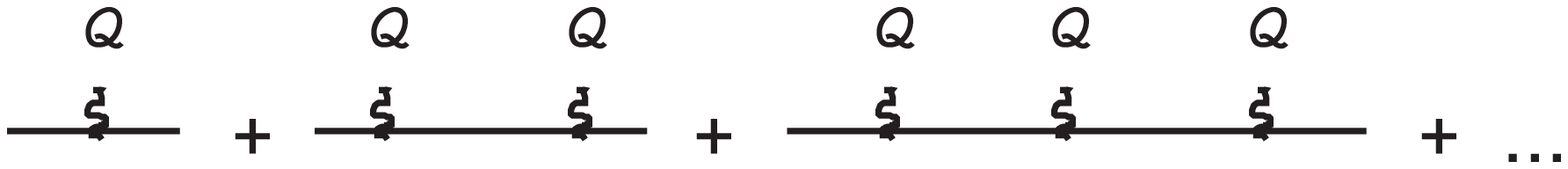}
\caption{ Examples of graphs associated with the expression (\protect\ref%
{rho3a}).}
\label{fig1_bhr}
\end{figure}
%%%%%%%%%%%%%%%%%%%%%%%%%%%%%%%%%%%%%%%%%%%%%%%%%%%%%%%%%%%%%%%%%%%%%%%%%%%%%
%
%
into Eq.~(\ref{rho2}), and replacement the order of integrations over the
fields $\Phi (\mathbf{r})$ and $Q(\mathbf{r})$ with the average over polymer
configurations $\mathbf{r}_{i}(s)$, yields the average over polymer
configurations in the form%
\begin{eqnarray}
&&\!\!\!\!\!\!\left\langle \rho _{\alpha }(\mathbf{r})\right\rangle  \notag
\\
&&\!\!\!=\frac{\int D\Phi (\mathbf{r})\Phi _{\alpha }(\mathbf{r})e^{-F_{%
\mathrm{int}}}\int DQ(\mathbf{r})e^{iQ\cdot \Phi }\left\langle e^{-iQ\cdot
\rho }\right\rangle _{0}}{\int D\Phi (\mathbf{r})e^{-F_{\mathrm{int}}}\int
DQ(\mathbf{r})e^{iQ\cdot \Phi }\left\langle e^{-iQ\cdot \rho }\right\rangle
_{0}},  \notag \\
&&  \label{rho2a}
\end{eqnarray}%
where $Q\cdot \rho $ stands for $\int d^{3}r\,Q_{\alpha }(\mathbf{r})\rho
_{\alpha }(\mathbf{r})$, and the brackets $\left\langle ...\right\rangle
_{0} $ means the average over conformations of ideal polymer chains
according to
\begin{equation}
\left\langle e^{-iQ\cdot \rho }\right\rangle _{0}=\int Dr_{i}(s)e^{-iQ\cdot
\rho }e^{-F_{\mathrm{el}}}.  \label{rho3}
\end{equation}%
To perform the average over polymer configurations we expand the first
exponent in expression (\ref{rho3}) in Taylor series. The mean value (\ref%
{rho3}) decomposes as products of averages over single polymer chains, which
have the structure
\begin{equation}
\!\int d^{3}r_{1}...\int d^{3}r_{k}Q_{\alpha }(\mathbf{r}_{1})...Q_{\alpha }(%
\mathbf{r}_{k})\left\langle \rho _{\alpha }(\mathbf{r}_{1})...\rho _{\alpha
}(\mathbf{r}_{k})\right\rangle ,  \label{rho3a}
\end{equation}%
where $k=0,1,...$, and $\alpha =A$, $B$. According to Ref.~\cite{stepanow95}
it is convenient to associate expression (\ref{rho3a}) with a graph
containing $k$ wavy legs, which are associated with $Q_{\alpha }(\mathbf{r}%
_{i})$. An example of graphs with $k=1,2,3$ is shown in Fig.~\ref{fig1_bhr}.
The continuous lines are associated with the propagator (\ref{bg}) for a
polymer blend in bulk and (\ref{G_diri}) for a polymer blend in the presence
of a hard wall, respectively. Consequently, the series (\ref{rho3}) is
associated with a product of $n_{A}$ lines for A polymers and $n_{B}$ lines
for B polymers containing an arbitrary number of wavy legs in each line.
Note that below we will consider the mean value (\ref{rho3}) in the
thermodynamic limit $n_{A}\rightarrow \infty $, $n_{B}\rightarrow \infty $, $%
\mathcal{V}\rightarrow \infty $ under the condition that the monomer
densities computed using the one-polymer Green's function are constant:
\begin{equation}
\left\langle \rho _{\alpha }(\mathbf{r})\right\rangle _{0}=N_{\alpha
}n_{\alpha }/\mathcal{V}\equiv \rho _{\alpha }.  \label{e1}
\end{equation}%
The corresponding density-density correlator reads
\begin{equation*}
\!\!\!\!\!\!\!\!\left\langle \rho _{\alpha }(\mathbf{r}_{2})\rho _{\alpha }(%
\mathbf{r}_{1})\right\rangle _{0}=\frac{1}{2}\rho _{\alpha }S_{\alpha \alpha
}(r_{1}-r_{2}).
\end{equation*}%
It is convenient to introduce the diagonal matrix $S_{\alpha \beta }$ such
that the Fourier transforms of the diagonal elements, $S_{\alpha \alpha }(k)$%
, are the bulk structure factor of the $\alpha $th component, which are
given by%
\begin{equation}
S_{\alpha \alpha }(k)=\rho _{\alpha }N_{\alpha }g(k^{2}R_{\mathrm{g},\alpha
}^{2}),  \label{str_f}
\end{equation}%
with $g(y)=2/y^{2}[\exp (-y)+y-1]$ being the Debye function.

In order to carry out integrations over the two-component field $Q(\mathbf{r}%
)$ in treating a blend in bulk \cite{stepanow95} one performs a partial
summation of the series (\ref{rho3}) (the latter can be carried out only in
the thermodynamic limit) by taking into account only the lines with one and
two insertions (wavy legs in Fig.~\ref{fig1_bhr}) in one polymer line. As a
result one obtains the expression
\begin{equation}
\exp \left[ iQ\cdot (\Phi -\left\langle \rho \right\rangle _{0})-\frac{1}{2}%
Q\cdot S\cdot Q\right] ,  \label{genF}
\end{equation}%
where $\left\langle \rho _{\alpha }\right\rangle _{0}$ is the average
monomer density (\ref{e1}). The integrations over $Q$ for a polymer blend in
the bulk is easily performed in Fourier space and result in%
\begin{equation}
\exp \left(-\frac{1}{2}\delta \Phi \cdot S^{-1}\cdot \delta \Phi \right).  \label{genF1}
\end{equation}%
where $\delta \Phi (\mathbf{r})=\Phi (\mathbf{r})-\left\langle \rho
\right\rangle _{0}$. The expression obtained after performing integrations
over $Q(\mathbf{r})$ can be written as $\exp (-H\{\delta \Phi \})$ with $%
H\{\delta \Phi \}$ being the Ginzburg-Landau functional \cite%
{leibler80,stepanow95}.

According to Ref.~\cite{stepanow95} the functional integration over $\delta
\Phi $ in Eq.~(\ref{rho2}) yields the monomer density $\left\langle \rho
_{\alpha }(\mathbf{r})\right\rangle $ as a series, which can be associated
with Feynman graphs similar to those in the theory of polymer solutions in
good solvent (see, for example, Ref.~\cite{descloizeaux}) with the
difference that the bare potentials are replaced by the effective ones
\begin{equation}
V^{\mathrm{eff}}=(V^{-1}+S)^{-1}.  \label{V-eff}
\end{equation}%
The lowest-order corrections to the monomer density are depicted in Fig.~\ref%
{fig2_bhr}. The external lines in these graphs are associated with the
expression
\begin{equation}
V^{\mathrm{ext}}=(V+S^{-1})^{-1}S^{-1},  \label{V-ext}
\end{equation}%
which can be written in the form $V^{-1}V^{\mathrm{eff}}$. The continuous
lines denote the bare bulk one-polymer Green's functions (\ref{bg}).

%%%%%%%%%%%%%%%%%%%%%%%%%%%%%%%   Figure 2  %%%%%%%%%%%%%%%%%%%%%%%%%%%%%%%%
\begin{figure}[tbph]
\includegraphics[clip,scale=0.48]{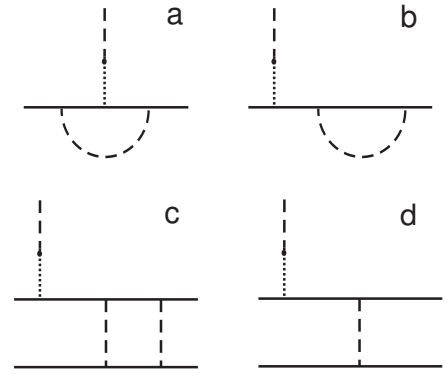}
\caption{ Examples of graphs contributing to the monomer concentration:
graphs $a$ and $b$ are first order and $c$ is second order in $V^{\mathrm{eff%
}}$. After renormalization the continuous line is associated with the
effective propagator. Graph $d$ with only one insertion of $V^{\mathrm{eff}}$
is identically zero after renormalization of internal lines.}
\label{fig2_bhr}
\end{figure}
%%%%%%%%%%%%%%%%%%%%%%%%%%%%%%%%%%%%%%%%%%%%%%%%%%%%%%%%%%%%%%%%%%%%%%%%%%%%%

\subsection{Screened effective potentials in an incompressible polymer blend}

\label{sec2c}

We now will consider in more details the properties of the effective
potential (\ref{V-eff}). We remind the reader that the quantities $S$, $V$,
and $V^{\mathrm{eff}}$ in Eq.~(\ref{V-eff}) are matrices. The elements of
the matrix $V^{\mathrm{eff}}$ are explicitly given by%
\begin{eqnarray}
V_{AA}^{\mathrm{eff}}(k) &=&\mathcal{R}(k)[-V+2V\chi S_{BB}+\chi ^{2}S_{BB}],
\notag \\
V_{AB}^{\mathrm{eff}}(k) &=&V_{BA}^{\mathrm{eff}}(k)=-\mathcal{R}(k)[V+\chi
],  \label{effp} \\
V_{BB}^{\mathrm{eff}}(k) &=&\mathcal{R}(k)[-V+2V\chi S_{AA}+\chi ^{2}S_{AA}],
\notag
\end{eqnarray}%
where for the sake of simplicity we have introduced the notation
\begin{eqnarray}
\mathcal{R}(k) &=&[-1-VS_{AA}-VS_{BB}  \notag \\
&+&2V\chi S_{AA}S_{BB}+\chi ^{2}S_{AA}S_{BB}]^{-1}.
\end{eqnarray}%
The behavior of the effective potentials in polymer blends was studied in
Refs.~\cite{brereton89,vilgis90,erukhimovich98}. In the following we will
consider an incompressible and athermic polymer blend, which in the
formalism under consideration is described in the limit $V\rightarrow \infty
$ and $\chi \rightarrow 0$. The effective potentials (\ref{effp}) simplify
in this limit to%
\begin{equation*}
V_{\alpha \beta }^{\mathrm{eff}}(k)=\frac{1}{S_{AA}+S_{BB}}.
\end{equation*}%
Using the explicit expression of the structure factor (\ref{str_f}) we
obtain for large polymer chains
\begin{equation}
V_{\alpha \beta }^{\mathrm{eff}}(k)=\frac{1}{12}\frac{1}{\rho
_{A}/l_{A}^{2}+\rho _{B}/l_{B}^{2}}k^{2}.  \label{V-eff1}
\end{equation}%
As it follows from Eq.~(\ref{V-eff1}) the expansion in powers of effective
potentials is in fact an expansion in inverse powers of the density.

We now will consider in more details the properties of the external
potentials (\ref{V-ext}) associated with external lines in graphs a, b, and c
in Fig.~\ref{fig2_bhr}, which are explicitly given by
\begin{eqnarray}
V_{AA}^{\mathrm{ext}}(k) &=&-\mathcal{R}(k)[1+VS_{BB}],  \notag \\
V_{AB}^{\mathrm{ext}}(k) &=&\mathcal{R}(k)[S_{AA}(V+\chi )],  \notag \\
V_{BA}^{\mathrm{ext}}(k) &=&\mathcal{R}(k)[S_{BB}(V+\chi )],  \notag \\
V_{BB}^{\mathrm{ext}}(k) &=&-\mathcal{R}(k)[1+VS_{AA}].  \notag
\end{eqnarray}%
In the case of athermic polymer blends the following identities hold
\begin{equation}
V_{AA}^{\mathrm{ext}}(k)-V_{AB}^{\mathrm{ext}}(k)=1,V_{BB}^{\mathrm{ext}%
}(k)-V_{BA}^{\mathrm{ext}}(k)=1.  \label{V-ext1}
\end{equation}%
For incompressible and athermic polymer blends $V_{\alpha \beta }^{\mathrm{%
ext}}(k)$ simplify to
\begin{equation*}
V_{AA}^{\mathrm{ext}}(k)=\frac{S_{BB}}{S_{AA}+S_{BB}},V_{AB}^{\mathrm{ext}%
}(k)=-\frac{S_{AA}}{S_{AA}+S_{BB}}.
\end{equation*}%
For large polymer chains we finally get
\begin{equation*}
V_{AA}^{\mathrm{ext}}(k)=\frac{\rho _{B}/l_{B}^{2}}{\rho _{A}/l_{A}^{2}+\rho
_{B}/l_{B}^{2}},V_{AB}^{\mathrm{ext}}(k)=-\frac{\rho _{A}/l_{A}^{2}}{\rho
_{A}/l_{A}^{2}+\rho _{B}/l_{B}^{2}},
\end{equation*}%
and similar for $V_{BA}^{\mathrm{ext}}(k)$ and $V_{BB}^{\mathrm{ext}}(k)$.
Note that in this limit the external lines are independent of the wave
number $\mathbf{k}$. Therefore, in the real space the external potential are
local and are given by the Dirac's $\delta$-function in this limit.

\subsection{Collective description of the polymer mixture in the presence of
a hard wall}

\label{sec2d}

We now will consider the collective description of a polymer blend in the
presence of a hard wall. In contrast to the collective description of
polymer blend in the bulk outlined in Sec.~\ref{sec2b} we have to use now
instead of the free propagator given by Eq.~(\ref{bg}) a propagator
fulfilling an appropriate boundary condition. In a theory based on the
statistical-mechanical description of single polymer chains, the boundary
conditions should coincide with those of single polymers, i.e., be the
Dirichlet boundary condition (\ref{dirichlet}). Since the behavior of a
polymer chain in solution and in a polymer melt in the presence of a wall
may be quite different, one can expect that the one-polymer Green's
functions in solution and in melt may obey different boundary conditions. A
consistent statistical-mechanical theory of polymer melt should be able, in
principle, to derive the boundary condition for one-polymer Green's function
appropriate for melt. We will show here that using a partial summation of
graphs it is possible to reformulate the description of polymer blend in
terms of the effective one-polymer Green's function. We will bring forward
the arguments that the latter should obey the reflecting boundary conditions.

In order to introduce the collective description in the presence of the
wall, we perform the same steps as in the bulk and arrive at the expression (%
\ref{rho2a}), and expand further $\left\langle \exp (-iQ\cdot \rho
)\right\rangle _{0}$ in Taylor series as given in Eq.~(\ref{rho3a}). In
contrast to the bulk the continuous lines are associated with bare
one-polymer Green's functions obeying the Dirichlet boundary condition (\ref%
{dirichlet}). The field $\Phi _{\alpha }(\mathbf{r})$ as well as $Q_{\alpha
}(\mathbf{r})$ are defined in the whole space as in the bulk formalism. The
mean density obtained by using the adsorbing boundary conditions is given by
Eq.~(\ref{rho4}) multiplied with the factor $n_{\alpha }/\mathcal{V}$. The
computation of the density-density correlation function $\left\langle \rho
_{\alpha }(r_{1})\rho _{\alpha }(r_{2})\right\rangle _{0}$ (no summation
over $\alpha $) for a polymer blend in the presence of a hard wall gives
%
%%%%%%%%%%%%%%%%%%%%%%%%%%%%%%%   Figure 3  %%%%%%%%%%%%%%%%%%%%%%%%%%%%%%%%
\begin{figure}[tbph]
\includegraphics[clip,scale=0.5]{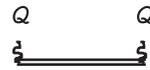}
\caption{Vertex with two insertions generated by the second term of Eq.~(%
\protect\ref{rho7}).}
\label{fig3_bhr}
\end{figure}
%%%%%%%%%%%%%%%%%%%%%%%%%%%%%%%%%%%%%%%%%%%%%%%%%%%%%%%%%%%%%%%%%%%%%%%%%%%%
%
\begin{eqnarray}
&&\!\!\!\!\!\!\!\!\left\langle \rho _{\alpha }(\mathbf{r}_{2})\rho _{\alpha
}(\mathbf{r}_{1})\right\rangle _{0}  \notag \\
&=&\frac{1}{2}\rho _{\alpha
}\int_{0}^{N}ds_{2}\int_{0}^{s_{2}}ds_{1}\left\langle \delta
[z_{2}-z(s_{2})]\delta [z_{1}-z(s_{1})]\right.   \notag \\
&&\times \left. \delta [\mathbf{r}_{2,\shortparallel }-\mathbf{r}%
_{2,\shortparallel }(s_{2})]\delta [\mathbf{r}_{1,\shortparallel }-\mathbf{r}%
_{2,\shortparallel }(s_{1})]\right\rangle _{0}  \notag \\
&=&\frac{1}{2}\rho _{\alpha }\int_{k_{\shortparallel }}\exp
(ik_{\shortparallel }(\mathbf{r}_{2,\shortparallel }-\mathbf{r}%
_{1,\shortparallel }))S_{\alpha \alpha }(z_{2},z_{1},k_{\shortparallel },N),
\label{rho5}
\end{eqnarray}%
where the Laplace transform of the diagonal element of the structure factor
is given by
\begin{equation}
S_{\alpha \alpha }(z_{2},z_{1},k_{\shortparallel },p)=\frac{2}{p^{2}}%
[G_{0}(z_{2}-z_{1},p+x)-G_{0}(z_{2}+z_{1},p+x)],  \label{rho6}
\end{equation}%
and where the notation $x=k_{\shortparallel }^{2}a^{2}$ has been introduced.
The nondiagonal elements of the matrix $S_{\alpha \beta }$ are zero. Note
that mean values of density products are zero if one of $z_{i}$ is negative,
so that the expression (\ref{rho6}) applies only for positive $z_{1}$ and $%
z_{2}$.

For a polymer blend in the presence of a wall the translationally invariant
part of the structure factor (\ref{rho6}) is defined only in the half space,
so that the integration over $Q$, which requires the inversion of $S_{\alpha
\beta }$, is not so straightforward. In this case we separate the
density-density correlator in two parts according to
\begin{eqnarray}
&&\left\langle \rho _{\alpha }(\mathbf{r}_{2})\rho _{\alpha }(\mathbf{r}%
_{1})\right\rangle _{0}=\left\langle \rho _{\alpha }(\mathbf{r}_{2})\rho
_{\alpha }(\mathbf{r}_{1})\right\rangle _{0\mathrm{b}}+\left\langle \rho
_{\alpha }(\mathbf{r}_{2})\rho _{\alpha }(\mathbf{r}_{1})\right\rangle _{0}
\notag \\
&&\hspace{9mm}-\left\langle \rho _{\alpha }(\mathbf{r}_{2})\rho _{\alpha }(%
\mathbf{r}_{1})\right\rangle _{0\mathrm{b}}  \notag \\
&&\hspace{9mm}=\left\langle \rho _{\alpha }(\mathbf{r}_{2})\rho _{\alpha }(%
\mathbf{r}_{1})\right\rangle _{\mathrm{0b}}+\left\langle \rho _{\alpha }(%
\mathbf{r}_{2})\rho _{\alpha }(\mathbf{r}_{1})\right\rangle _{0\mathrm{s}},
\label{rho7}
\end{eqnarray}%
and perform a partial summation by taking into account in every line only
the first term in Eq.~(\ref{rho7}). In proceeding in this way we fix the
reference state to be that of the bulk far from the wall. The prize to pay
is that the 2nd term in Eq.~(\ref{rho7}) has to be taken into account as a
vertex with two insertions, which is shown in Fig.~\ref{fig3_bhr}.

The summation over lines with one and two insertions in one line results
exactly in the expression given by Eq.~(\ref{genF}) with the average density
given now by Eq. (\ref{rho4}). The terms in the series (\ref{rho3}) with
more than two fields $Q(\mathbf{r})$ along one line [and two fields
corresponding to the 2nd term in Eq.~(\ref{rho7})] can be obtained from Eq.~(%
\ref{genF}) and consequently Eq.~(\ref{genF1}) as derivatives with respect
to $\delta \Phi (\mathbf{r})$. To compute the concentration profile
according to Eq.~(\ref{rho2}) one should perform integration over the field $%
\Phi (\mathbf{r})$. While after integrations over $Q$ the series (\ref{rho3}%
) depends on $\delta \Phi (\mathbf{r})=\Phi (\mathbf{r})-\left\langle \rho
\right\rangle $, the interaction part of the free energy (\ref{b21}) has the
form $F_{\mathrm{int}}=\frac{1}{2}\Phi \cdot V\cdot \Phi $. Rewriting the
latter in terms of $\delta \Phi $ yields%
\begin{equation}
F_{\mathrm{int}}=\frac{1}{2}\delta \Phi \cdot V\cdot \delta \Phi +\delta
\Phi \cdot V\cdot \left\langle \rho \right\rangle ,
\end{equation}%
where the linear term in $\delta \Phi $ has the same form as interaction
with an external field in the formalism of $\Phi ^{4}$ theory. For an
incompressible and athermic polymer blend, which we consider in the present
work, the linear term vanishes.

Similar to the consideration in bulk the expression obtained after
performing integrations over $Q(\mathbf{r})$ can be written as $\exp
(-H\{\delta \Phi \})$ with $H\{\delta \Phi \}$ being a Ginzburg-Landau
Hamiltonian including the surface terms. However, in contrast to the
effective surface Hamiltonian used in many studies \cite{binder-actap95} the
corresponding terms are not localized at the surface only \cite{unpb}. The
integration over the field $\Phi (r)$ can be performed in the same way as
for bulk.

The collective description developed above is based on the concept of the
effective potential, which takes into account the screening of
monomer-monomer interactions in a melt. However, the effect of the wall is
taken into account as in the case of diluted polymers via the Dirichlet
boundary condition for one-polymer Green's function and leads to an
inhomogeneous monomer density for distances up to the gyration radius.
However, in a melt the density is expected to be rather homogeneous at
distances $z<R_{g}$. This is the result of the interplay of the interaction
with the wall and the incompressibility of the polymer melt. While in the
polymer solution (which is a liquid and as such is incompressible) the
entropic repulsion with the wall favor the presence of solvent molecules at
the wall. In the case of the melt this is not anymore the case, because the
place of monomers being repulsed from the wall, will be occupied by monomers
belonging to another polymer which at that moment are not or less repulsed
from the wall. Due to this the melt density similar to the total density of
the solution will not tend to zero in approaching the surface. We expect
that the effect of the wall on the behavior in the polymer melt can be
formulated in terms of the renormalized one-polymer Green's function, which
should guarantee the uniformity of the density, and according to this
should obey a boundary condition, which is different from the Dirichlet
boundary condition. We now show that, indeed, the partial summation of
graphs including insertions into continuous lines enables one to formulate
the description of a polymer melt in terms of the effective one-polymer
Green's function. We will consider for simplicity the renormalization of the
bare one-polymer Green's functions in the expression for the concentration
profile
\begin{eqnarray}
&&\!\!\!\!\!\!\!\!\!\!\!\!\!\!\!\!\!\!\left\langle \rho _{\alpha
}(z)\right\rangle  \notag \\
&&\!\!\!\!\!\!\!\!\!\!\!\!\!\!\!\!\!\!=n_{\alpha }\int_{0}^{N_{\alpha
}}ds\int\limits_{0}^{\infty }dz^{\prime }G(z^{\prime },z,N_{\alpha
}-s)\int\limits_{0}^{\infty }dz^{\prime \prime }G(z,z^{\prime \prime },s),
\label{rho8}
\end{eqnarray}%
where the bare one-polymer propagator obeys the Dirichlet boundary
condition. The graphs $b$ and $c$ in Fig.~\ref{fig2_bhr} contribute to the bare
Green's functions in (\ref{rho8}). Using property (\ref{V-ext1}) and
expressing the external potentials $V_{AA}^{\mathrm{ext}}$ as $1+V_{AB}^{%
\mathrm{ext}}$ we can divide this contribution into two parts. The first one
is given by graphs $b$ and $c$, in which the external line associated with $%
1 $ and which renormalizes the one-polymer propagators in (\ref{rho8}). The
%
%%%%%%%%%%%%%%%%%%%%%%%%%%%%%%%%   Figure 4  %%%%%%%%%%%%%%%%%%%%%%%%%%%%%%%%
\begin{figure}[tbph]
\includegraphics[clip,scale=0.5]{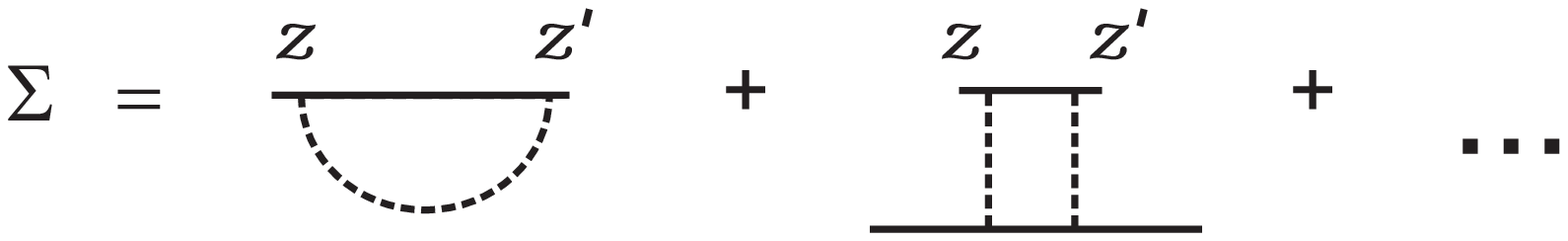}
\caption{The lowest-order graphs contributing to the self-energy. The
continuous lines are associated with the effective one-polymer propagator $%
G_{r,\protect\alpha }$. }
\label{fig4_bhr}
\end{figure}
%%%%%%%%%%%%%%%%%%%%%%%%%%%%%%%%%%%%%%%%%%%%%%%%%%%%%%%%%%%%%%%%%%%%%%%%%%%%
%
second part, with the external line associated with $V_{AB}^{\mathrm{ext}}$,
together with the graph $a$ describes the fluctuations corrections to the
concentration profile. The renormalization to the first order can be
extended to higher orders, with the result that the bare continuous lines
will be replaced by the effective ones associated with the effective
one-polymer Green's function. This procedure corresponds to reduction of the
whole set of graphs to the skeleton graphs, i.e. the graphs without
insertions into internal lines. The only exception are the graphs $b$ and $c$ in
Fig.~\ref{fig2_bhr}, which are due to the recasting of $V_{AA}^{\mathrm{ext}%
} $. The renormalization of one-polymer graphs due to insertions into the
internal lines can be represented using the Dyson equation%
\begin{equation}
G_{r}^{-1}=G^{-1}-\Sigma ,  \label{dyson}
\end{equation}%
where $\Sigma $ is the self-energy, which takes into account insertions
along the chain. Note that $G$, etc., in Eq.~(\ref{dyson}) are matrices with
respect to spatial coordinates. Examples of graphs contributing to $\Sigma $
are given in Fig.~\ref{fig4_bhr}.

\begin{figure}[tbph]
\includegraphics[clip,scale=0.18]{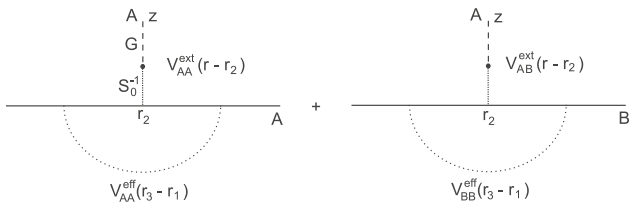}
\caption{The Feynman diagrams giving the leading contribution to the excess
monomer concentration.}
\label{fig5_bhr}
\end{figure}

As a result of the partial summation of graphs taking into account the
insertions into internal lines according to Eq.~(\ref{dyson}) the
lowest-order contribution to the density profile (\ref{rho8}) changes to%
\begin{eqnarray}
\left\langle \rho _{\alpha }(z)\right\rangle &=& n_{\alpha
}\int_{0}^{N_{\alpha }}ds\int\limits_{0}^{\infty }dz^{\prime }G_{r,\alpha
}(z^{\prime },z,N_{\alpha }-s)  \notag \\
&\times& \int\limits_{0}^{\infty }dz^{\prime \prime }G_{r,\alpha
}(z,z^{\prime \prime },s).  \label{rho9}
\end{eqnarray}%
The fluctuation corrections to Eq.~(\ref{rho9}) are given by the skeleton
graphs in Figs.~\ref{fig2_bhr} and \ref{fig5_bhr}. As a result of the
partial summation the bare one-polymer propagators $G$ are replaced by the
effective ones $G_{r,\alpha }$.

Equation (\ref{dyson}) with $\Sigma $ given as an infinite set of graphs is
the basis of the self-consistent computation of the effective one-polymer
Green's function in the polymer blend under presence of a hard wall. The
solution of this equation is a difficult task which goes beyond the scope of
the present article.

%%%%%%%%%%%%%%%%%%%%%%%%%%%%%%%   Figure 5  %%%%%%%%%%%%%%%%%%%%%%%%%%%%%%%%
%%%%%%%%%%%%%%%%%%%%%%%%%%%%%%%%%%%%%%%%%%%%%%%%%%%%%%%%%%%%%%%%%%%%%%%%%%%%

Fortunately, the form of $G_{r,\alpha }$ in polymer fluid can be found from
general arguments avoiding the direct solution of Eq.~(\ref{dyson}).
According to the above discussion we expect that the density profile in an
incompressible fluid in the presence of neutral wall will be uniform. On the
another hand the density profile without taking into account the
fluctuations is given by Eq.~(\ref{rho9}). As we have shown in Sec.~\ref%
{sec2a} the computation of the density using the one-polymer Green's
function obeying the reflecting boundary condition gives a homogeneous
density. Due to this we identify the effective one-polymer propagators $%
G_{r,\alpha }$ with that obeying the reflecting boundary condition.
Deviations from Silberberg hypothesis in thin polymer films were studied
recently in Refs.~\cite{semenov-johner03,cavallo05,rehse01,zhao93}.

Figure~\ref{fig4_bhr} shows that the contributions to the self-energy does not
reduce to an effective potential as it was assumed in Ref.~\cite%
{wu-fredrik-carton} in the approach based on the self-consistent field
theory. While the first graph in Fig.~\ref{fig4_bhr} takes into account the
monomer-monomer interactions along one polymer, the 2nd graph (and
higher-order graphs) takes into account monomer-monomer interactions
between different polymers. Since the graphs contributing to the self-energy
take into account the many particle interactions characteristic for a melt,
we expect that the effective one-polymer Green's function $G_{r,\alpha }$
obeys the boundary conditions appropriate for an incompressible liquid, i.e.,
the reflecting boundary conditions. This conclusion is supported by the
following argument. In polymer melts similar to semidilute polymer
solutions the relevant quantity governing the properties of the system is
the number of monomers between two subsequent cross-links along the chain,
which for polymer melts is of order of unity, instead of the chain length $N$%
. Consequently, in the polymer melt the effect of the wall on the monomers,
which are close to the wall, will be similar to that on solvent molecules in
solution. However, the monomers are classical objects, which are described
in the relaxational regime. For a single monomer (or a solvent molecule),
which dynamics is described by the Langevin equation, the steady-state
distribution function is given by Boltzmann distribution $\exp
[-U(z)/k_{B}T] $ and is therefore constant in the space between two walls.
This makes clear that the Dirichlet boundary condition is irrelevant in
dense polymeric systems.

It is well known that a polymer configuration corresponds to the trajectory
of a quantum particle for imaginary times. According to this the problem of
boundary condition in polymer melts is expected to have its counterpart in
quantum fluids in the presence of a neutral boundary. While the wave
functions of single particles obey the Dirichlet boundary condition at the
wall, the density of the fluid is not required to be zero at the wall~\cite%
{khalatnikov}.

\section{Computation of the excess monomer concentration}

\label{comput}

The skeleton graphs in Fig.~\ref{fig2_bhr} give the fluctuational part to
the density profile. The free end of the external line is associated with
the argument of the monomer density $z$ (due to the symmetry along the wall
the monomer density does not depend on $\mathbf{r}_{\shortparallel }$). The
explicit calculation shows that the one-loop graphs, where the external line
is located outside the loop (graphs $b$ and $c$ in Fig.~\ref{fig2_bhr}), are
negligible for large $N$. The leading contribution is due to the graph $a $
and the related graph, which describes the effect of B polymers on the
concentration of A polymers. These graphs are shown in more details in Fig.~%
\ref{fig5_bhr}.

We now will consider the computation of the concentration of say the
component A in the presence of a hard wall. We will assume that the
statistical segment length of the polymer A is larger than that of the
polymer B, $l_{A}>l_{B}$, so that the polymer A is stiffer. The contribution
to the excess concentration to the lowest order in powers of the effective
potentials is given by graphs in Fig.~\ref{fig5_bhr}. To conduct
calculations it is convenient to consider the Laplace transform with respect
to the contour length $N$. The analytical expression associated with the
first graph in Fig.~\ref{fig5_bhr} is given by
\begin{eqnarray}
&&\!\!\!\!\!\!-\frac{\rho _{A}V_{AA}^{\mathrm{ext}}}{8\pi ^{3}N_{A}}\int
d^{2}q_{\shortparallel }\int dq\ V_{AA}^{\mathrm{eff}}(q{_{\shortparallel
}^{2}+}q^{2})  \notag \\
&&\!\!\!\!\!\!\times \frac{q^{2}a^{2}e^{-2z\sqrt{p+x}/a}+2qa\sqrt{p+x}\sin
(qz)e^{-z\sqrt{p+x}/a}+p{+}x}{p^{2}(p+x)(p+x+q^{2}a^{2})^{2}},  \notag \\
&&  \label{md1}
\end{eqnarray}%
where $p$ is Laplace conjugate to $N$ and $x=q_{\shortparallel }^{2}a^{2}$.
The analytical expression of the 2nd graph in Fig.~\ref{fig5_bhr} is
obtained from Eq.~(\ref{md1}) using the replacements%
\begin{equation*}
V_{AA}^{\mathrm{ext}}\rightarrow V_{AB}^{\mathrm{ext}},\ V_{AA}^{\mathrm{eff}%
}\rightarrow V_{BB}^{\mathrm{eff}},\ \rho _{A}\rightarrow \rho _{B},\
N_{A}\rightarrow N_{B}.
\end{equation*}%
Note that the factor $-1$ is due to the fact that $V$ and $V^{\mathrm{eff}}$
appear with the sign minus in the exponential of the statistical weight of
polymer configurations. The $k^{2}$ dependence of the effective potentials (%
\ref{V-eff1}) leads to the divergence of the integrals over the wave vector
in Eq.~(\ref{md1}) at the upper limit of integration. However, the effective
potentials acquire for finite $V$ their bare values for large $k$, so that
the integral converges at the upper limit of the integration. Therefore, for
finite $V$ the effective potentials are screened only for lengths larger
than the local length
\begin{equation*}
l_{c}\approx V^{-1/2}(\rho _{A}{/}l{_{A}^{2}+}\rho _{B}{/}l_{B}^{2})^{-1/2},
\end{equation*}%
which is obtained from the explicit expressions of the effective potentials (%
\ref{effp}). The derivation shows that this length is the same for both
polymers. We expect that for finite $V$ the polymer blend can be considered
as an incompressible only for lengths larger than $l_{c}$. In order to
simplify the integration over the wave vector in Eq.~(\ref{md1}) we use the
athermic and incompressible limit of the effective potentials (\ref{V-eff1}%
), but restrict the integration to wave vectors smaller than the cutoff
value $\Lambda \simeq l_{c}^{-1}$.

The inspection of Eq.~(\ref{md1}) shows that it (and the expression
associated with graphs with the external line being outside the loop)
contains a $z$ independent contribution to the excess concentration of the
density. The straightforward computation yields the renormalization of the
bulk monomer concentration as
\begin{equation}
\tilde{\rho}_{A}=\rho _{A}\left( 1+\left( 1-2\right) \frac{3\rho _{B}\Lambda
}{4\pi }\frac{l_{A}^{2}l_{B}^{2}}{(\rho _{A}l_{B}^{2}+\rho _{B}l_{A}^{2})^{2}%
}(\frac{1}{l_{B}^{2}}-\frac{1}{l_{A}^{2}})\right) .  \label{md2}
\end{equation}%
The factor $2$ in Eq.~(\ref{md2}) accounts for graphs similar to the graphs $%
b$ and $c$ in Fig.~\ref{fig2_bhr} but with the external lines being on the
right side of the interaction line. Note that the mass divergences \cite%
{descloizeaux} of the graphs $b$ and $c$\ are omitted, that implies the
regularization of expression (\ref{rho1}) with respect to the mass
divergences at the beginning. Equation (\ref{md2}) shows that even in the
bulk the packing effects change the bare density of the constituents: the
concentration of the stiffer polymer becomes smaller. Without incorporating
the possibility for a local nematic ordering, which is not taken into
account in the model of a Gaussian polymer chain, polymers with larger
statistical segment length are expected to have smaller density. Note that
the renormalization of the bulk composition is local, and the comparison of
Eq.~(\ref{md2}) with the corresponding expression for $\tilde{\rho}_{B}$
shows that the total density of the blend does not change. Although the
renormalization of the bulk composition given by Eq.~(\ref{md2}) is somewhat
unexpected, its necessity can be explained qualitatively as follows. The
density of an incompressible liquid at given $T$ and $\mathcal{V}$ is
determined by interactions between the molecules, and cannot be chosen
arbitrarily as in gas-like systems. Thus, in application of the
coarse-grained model under consideration to polymer blend Eq. (\ref{md2})
describes the renormalization of bare concentrations towards their
concentrations in the polymer melt, which are determined by monomer-monomer
interactions.

The $z$ dependent part of Eq.~(\ref{md1}) gives the excess monomer
concentration as a function of the distance to the wall. The integration
over the wave vector yields the simple expression%
\begin{eqnarray}
&&-\frac{V_{0}}{8a^{5}p^{2}\pi z}\left[ e^{-({2z}/{a})\sqrt{p}}(a-z\sqrt{p}%
)\right.  \notag \\
&&\left. -e^{-({2z}/{a})\sqrt{p+a^{2}\Lambda ^{2}}}(a-\frac{zp}{\sqrt{%
p+a^{2}\Lambda ^{2}}})\right]  \notag \\
&&-\frac{V_{0}\Lambda }{2a^{4}\pi ^{2}p^{2}}\left( \Gamma _{0}\left( \frac{2z%
}{a}\sqrt{p}\right) -\Gamma _{0}\left[ \frac{2z}{a}\sqrt{p+a^{2}\Lambda ^{2}}%
\right] \right) ,\ \ \ \ \ \   \label{md3}
\end{eqnarray}%
where $\Gamma _{\alpha }(x)=\int_{x}^{\infty }dt\ t^{\alpha -1}\exp (-t)$ is
the incomplete Gamma function, and the notation
\begin{equation*}
V_{0}=\frac{1}{12}\frac{\rho _{A}\rho _{B}}{N_{A}l_{B}^{2}}\frac{1}{(\rho
_{A}/l_{A}^{2}+\rho _{B}/l_{B}^{2})^{2}}
\end{equation*}%
is introduced. To obtain the excess density one should add to expression (%
\ref{md3}), which is associated with the first graph in Fig.~\ref{fig5_bhr},
the corresponding expression associated with the 2nd graph in Fig.~\ref%
{fig5_bhr}.

We will first compute the excess concentration of the stiffer (A) polymer at
the surface $\delta \rho _{A}(z=0)$. To that end we put $z=0$ in Eq.~(\ref%
{md3}), take into account the second graph in Fig.~\ref{fig5_bhr}, and
perform the inverse Laplace transform. For large $N$ we obtain the result
\begin{eqnarray}
&&\delta \rho _{A}(z=0)=\frac{3}{4\pi ^{2}}\Lambda \frac{\rho _{A}\rho
_{B}l_{A}^{2}l_{B}^{2}}{(\rho _{A}l_{B}^{2}+\rho _{B}l_{A}^{2})^{2}}   \notag
\\
&&  \hspace{10mm}\times \left[ \frac{1}{l_{B}^{2}}\ln (a_{B}^{2}\Lambda
^{2}N_{B})-\frac{1}{l_{A}^{2}}\ln (a_{A}^{2}\Lambda ^{2}N_{A})\right].  \label{md4}
\end{eqnarray}%
If both polymers have the same gyration radius $R_{g}=a\sqrt{N}$ Eq.~(\ref%
{md4}) simplifies to
\begin{equation}
\delta \rho _{A}(z=0)=\frac{3}{4\pi ^{2}}\Lambda \frac{\rho _{A}\rho
_{B}l_{A}^{2}l_{B}^{2}}{(\rho _{A}l_{B}^{2}+\rho _{B}l_{A}^{2})^{2}}\left(\frac{1%
}{l_{B}^{2}}-\frac{1}{l_{A}^{2}}\right){\ln (}\Lambda ^{2}R_{g}^{2}{)}.
\label{md5}
\end{equation}%
The excess concentration at the wall for polymer blend differing only in
degrees of polymerization is derived from Eq.~(\ref{md4}) as
\begin{equation*}
\delta \rho _{A}(z=0)=\frac{3}{4\pi ^{2}}\frac{\Lambda }{l^{2}}\frac{\rho
_{A}\rho _{B}}{(\rho _{A}+\rho _{B})^{2}}\ln \frac{N_{B}}{N_{A}}.
\end{equation*}

%%%%%%%%%%%%%%%%%%%%%%%%%%%%%%%   Figure 6  %%%%%%%%%%%%%%%%%%%%%%%%%%%%%%%%
\begin{figure}[tbph]
\includegraphics[clip,scale=0.6]{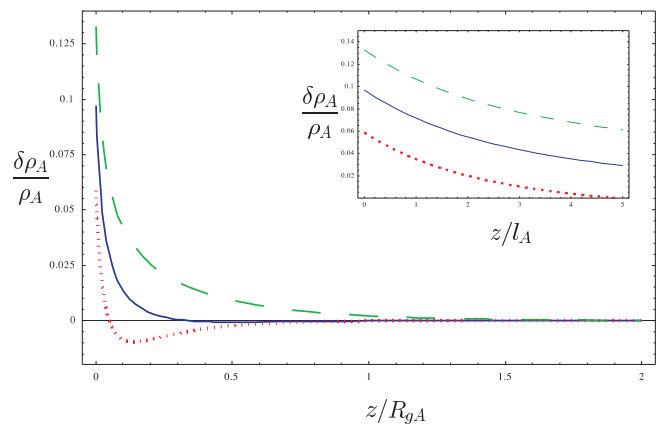}
\caption{(Color online) Concentration profile of A polymers as a function of
the distance to the surface for different values of $N_{B}$, and $l_{A}=1.5$%
, $l_{B}=1$, $\Lambda ^{-1}=1.55$, $\protect\rho _{A}=\protect\rho _{B}=0.5$%
. The continuous line: $N_{A}=N_{B}=10^{4}$; dashes: $N_{B}=5\times 10^{4}$;
dots: $N_{B}=2\times 10^{3}$. The inset shows the concentration profile in
the vicinity of the surface as a function of the distance measured in units
of $l_{A}$.}
\label{fig6_bhr}
\end{figure}
%%%%%%%%%%%%%%%%%%%%%%%%%%%%%%%%%%%%%%%%%%%%%%%%%%%%%%%%%%%%%%%%%%%%%%%%%%%%
The latter shows that the shorter polymers are present in excess at the
wall. Notice that the excess concentration depends logarithmically on the
number of segments $N$. The contribution to the excess concentration at $z=0$
associated with graphs $b$ and $c$ in Fig.~\ref{fig2_bhr} reads
\begin{eqnarray}
&&\delta \rho _{A}(z=0)=-\frac{1}{8(6\pi )^{3/2}l_{A}^{2}}\frac{\rho _{A}}{%
(\rho _{A}l_{B}^{2}+\rho _{B}l_{A}^{2})^{2}}  \notag \\
&&\hspace{3mm}\times \lbrack \frac{l_{B}^{5}\rho _{A}}{\sqrt{N_{A}}}\ln (%
\frac{4}{9}a_{A}^{2}\Lambda ^{2}N_{A}{)}+\frac{l_{A}^{5}\rho _{B}}{\sqrt{%
N_{B}}}\ln (\frac{4}{9}a_{B}^{2}\Lambda ^{2}N_{B}{)}].\ \ \ \ \   \label{md6}
\end{eqnarray}%
Due to the factor $N^{-1/2}$ the latter vanishes for large $N$. Note that
for conformationally asymmetric polymers of the same gyration radius the
sign of Eq.~(\ref{md6}) is opposite to that of Eq.~(\ref{md5}). The increase
of $\delta \rho _{A}(z=0)$ with $N$ agrees qualitatively with the results of
numerical simulations and calculations using the integral equation
theory~\cite{yethiraj94}.

\begin{figure}[tbph]
\includegraphics[clip,scale=0.6]{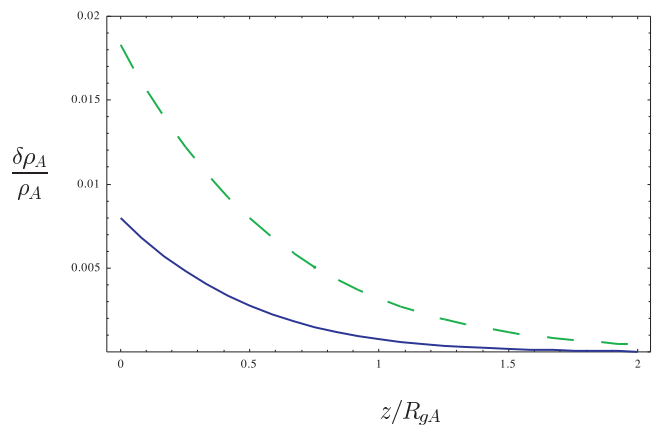}
\caption{(Color online) Concentration profile of A polymers as a function of
the distance to the surface for different values of $N_{B}$, and $%
l_{A}=l_{B}=1.5$, $N_{A}=10^{4}$, $\Lambda ^{-1}=1.55$, $\protect\rho _{A}=%
\protect\rho _{B}=0.5 $. The continuous line: $N_{B}=2\times 10^{4}$;
dashes: $N_{B}=5\times 10^{4} $. }
\label{fig7_bhr}
\end{figure}

To compute $\delta \rho _{A}(z)$ for arbitrary $z$ one should perform the
inverse Laplace transform of Eq.~(\ref{md3}). Since it cannot be performed
analytically, we have used a numerical routine (Durbin) for inverse Laplace
transform in Mathematica. The results of the numerical calculation of the
excess concentration of stiffer polymers $\delta \rho _{A}(z)$ for different
values of the degrees of polymerization of more flexible polymer are shown
in Fig.~\ref{fig6_bhr}. It shows that the increase of $N_{B}$ results in an
increase of the excess concentration of the A polymer. For $N_{B}<N_{A}$ the
concentration of A polymers is still in excess in the vicinity of the wall,
but becomes lower than in the bulk for intermediate distances, i.e., the
B polymers are in excess at these distances. These results are in agreement
with numerical simulations and computations using the integral equation
theory \cite{yethiraj94}. Figure~\ref{fig7_bhr} shows the result of the
computation of the excess concentration of the shorter polymers in a polymer
blend consisting of chemically identical polymers, which differ only in
their degrees of polymerization. Figure~\ref{fig7_bhr} shows that shorter
polymers are present in excess in the vicinity of the wall. This finding is
in qualitative agreement with the result predicted in Ref.~\cite{hariharan90}
and observed in Refs.~\cite{walton-mayes96} and \cite%
{schaub96,hong94,hopkinson95}.

%%%%%%%%%%%%%%%%%%%%%%%%%%%%%%%   Figure 7  %%%%%%%%%%%%%%%%%%%%%%%%%%%%%%%%
%%%%%%%%%%%%%%%%%%%%%%%%%%%%%%%%%%%%%%%%%%%%%%%%%%%%%%%%%%%%%%%%%%%%%%%%%%%%%

The excess of shorter polymers in the case under consideration is compatible
with the excess of the solvent at the wall in a polymer solution. The latter
corresponds to the limit, when the polymerization degree of shorter polymers
tends to unity. However, to describe this limit one has to take into account
the higher-order terms in the perturbation series for the concentration
profile.

Note that the both cases we have considered above ($l_{A}\neq l_{B}$, $%
R_{gA}=R_{gB}$ and $l_{A}=l_{B}$, $N_{A}\neq N_{B}$) follow from the general
formula (\ref{md3}).

We now will give a qualitative explanation of the different behavior of
polymers in the blend under the influence of a hard wall. A single polymer
in a dilute solution obeys the Dirichlet boundary condition. As a
consequence of the boundary condition the number of configurations available
to the polymer chain lowers with the decrease of the distance to the wall.
The latter results in an entropic repulsion of the polymer from the wall,
and is responsible for the vanishing of the density at the wall. According
to this the solvent molecules are favored in the vicinity of the wall with
respect to the polymer monomers. A simple calculation using the distribution
function obeying the Dirichlet boundary condition shows that the force
acting on the free end of the polymer at a given distance to the wall is
controlled by the gyration radius of the polymer $R_{g}=a\sqrt{N}$.

A completely different behavior takes place in the case of incompressible
polymer melts, where the entropic repulsion from the wall is balanced by the
melt pressure with the consequence that the density is uniform. However,
there is a difference in the behavior of the polymers in the vicinity of the
wall for melt composed of different polymers. We consider first a polymer
blend composed of polymers which differ only in degrees of polymerization.
In a layer with the thickness equal to the gyration radius of larger
polymers, the larger polymer experiences the entropic force from the wall
while the shorter polymer does not. Due to this the larger polymer increases
its distance to the wall, which will be occupied by shorter ones, in order
that the total density will remain constant. The asymmetry in the behavior
of polymers in the vicinity of the wall appears even in a polymer melt
composed of identical polymers. According to the above argument the monomers
of a polymer coil, which has contacts with the wall, are disfavored with
respect to the ends of polymer coils which do not have contacts with the
wall. Due to this the polymer ends are expected to be present in excess in
the vicinity of the wall. The effect of the distribution of polymer ends on
the surface tension was studied in Ref.~\cite{degennes88}. A quantitative
study of the distribution of polymer ends using the self-consistent field
theory was performed in Ref.~\cite{wuetal95}.

For polymers with different statistical segment lengths, but the same
gyration radius the difference in the behavior in the vicinity of the wall
can be explained qualitatively as follows. The monomer density of a polymer
coil is given by $\rho _{\mathrm{c}}=N/R_{g}^{3}=a^{-2}/R_{g}$, while the
surface density of a coil is $\rho _{\mathrm{s}}=\rho _{\mathrm{c}%
}R_{g}=a^{-2}$. Therefore, the surface density $\rho _{s}$ of the stiffer
polymer is smaller. It is likely to expect that the repulsive effect of the
wall on the coil is proportional to $\rho _{s}$. According to this the
repulsive effect of the wall is stronger for more flexible polymers. This is
the reason that the monomers of stiffer polymers will be favored in the
vicinity of the wall. The surface enrichment $\delta \rho _{A}$ is expected
to be proportional to the differences of surface densities, i.e., $\delta
\rho _{A}\sim \rho _{\mathrm{s}}^{B}-\rho _{\mathrm{s}}^{A}$, which agrees
with our quantitative result (\ref{md5}). According to this qualitative
consideration the difference in surface densities $\rho _{\mathrm{s}%
}^{B}-\rho _{\mathrm{s}}^{A}$ is a drive for the conformational asymmetry.
Since the monomers within the layer of thickness $R_{g}$ are affected by the
wall, we expect that the excess concentration will depend on $R_{g}$.
However, the logarithmic dependence on $R_{g}$ in Eq.~(\ref{md5}) is
difficult to derive using only the hand wavy arguments.

Note that in the above computation of the excess concentration $\delta \rho
_{A}(z)$ we have taken into account the lowest-order correction in the
series in powers of effective potentials. The effective potentials according
to Eq.~(\ref{V-eff1}) are inversely proportional to the density, so that the
perturbation expansion in powers of effective potentials is a series in
inverse powers of the density. However, since the polymer melt has a fixed
density, the inverse density is not a small parameter. The magnitude of the
first-order correction can be controlled by considering polymers having the
same gyration radius and small differences in $l_{A}$ and $l_{B}$, or
polymers with small differences in $N_{A}$ and $N_{B}$ for $l_{A}=l_{B}$.
However, it is not clear without explicit computations, if the 2nd order
term is smaller than the 1st order one under the above conditions. From the
general point of view one would expect the following bounds on the total
effect of the perturbation series. As already mentioned above for polymers
differing only in degrees of polymerization the effect of the whole
perturbation series should recover in the limit $N_{A}\ll N_{B}$ the
behavior in polymer solutions, where the polymer concentration will tend to
zero in approaching the surface. For polymers differing in flexibility the
concentration of the stiffer polymer at the wall cannot exceed the total
density of the polymer blend in bulk. In other words the concentration of
the more flexible polymer cannot be negative. This determines the upper
limit of applicability of our results given by Eqs.~(\ref{md4}) and (\ref%
{md5}).

\section{Conclusions}

\label{concl}

To summarize, we have generalized the Edwards' collective description of
dense polymer systems in terms of effective potentials to polymer blends in
the presence of a surface. Using this formalism we have studied an
incompressible athermic polymer blend of conformationally asymmetric
polymers, which differ in statistical segment lengths, in the presence of a
hard wall. We have computed the excess concentrations of constituents to the
first order in powers of effective potentials. We have found that stiffer
polymers are in excess in the vicinity of the surface, and that the
concentration excess at the surface depends logarithmically on the degrees
of polymerization. For polymer blends differing only in degrees of
polymerization the shorter polymers are in excess at the wall. Our results
are in agreement with numerical results available in the literature. The
present method can be applied in a straightforward way to study the behavior
of polymer blends and copolymer melt in the presence of selective surfaces,
to study the dimensions of polymer molecules in the melt, the distribution
of polymer ends, etc.

\begin{acknowledgments}
We would like to thank H. Angerman, and A. Johner for useful discussions. A
financial support from the Deutsche Forschungsgemeinschaft, SFB 418 is
gratefully acknowledged.
\end{acknowledgments}

\end{document}